\documentclass[aps,showpacs,prd,twocolumn,nofootinbib,nobibnotes]{revtex4-2}
\usepackage{graphicx}
\usepackage{amssymb}
\usepackage{amsmath}
\usepackage{color}
\usepackage{float}
\usepackage[normalem]{ulem}
\usepackage{accents}
\usepackage{graphicx}
\usepackage{graphicx}
\usepackage{amsfonts}
\usepackage[colorlinks=true,
pdfstartview=FitV,linkcolor=blue,
citecolor=blue,urlcolor=blue,breaklinks=true]
{hyperref}
\usepackage{array}
\usepackage{float}
\usepackage{placeins}
\usepackage[dvipsnames]{xcolor}
\usepackage{csquotes}
\usepackage{bbold}
\usepackage{units}
\usepackage{makecell}
\usepackage{enumitem}
\usepackage{dsfont}
\usepackage{upgreek}
\usepackage{booktabs}

\newcolumntype{C}[1]{>{\centering\arraybackslash}m{#1}}

\renewcommand{\eqref}[1]{\mbox{Eq.~(\ref{#1})}}

\definecolor{ForestGreen}{rgb}{0.13,0.55,0.13}

\usepackage{fixmath}

\newcommand{\orcid}[1]{\href{https://orcid.org/#1}{\includegraphics[width=10pt]{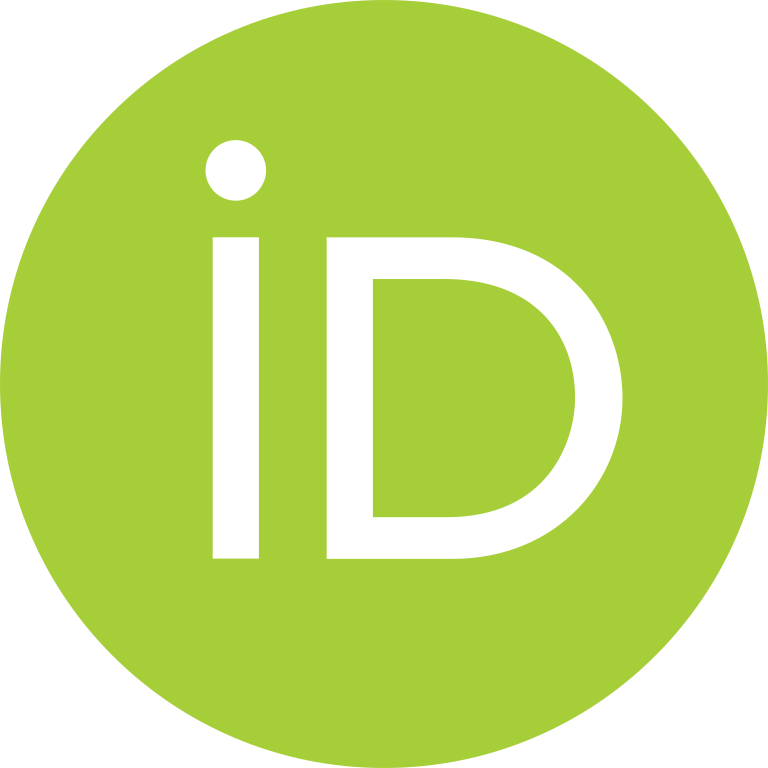}}}

\RequirePackage{color}

\begin{document}
	
	\title{Hypothesis of a bi-isotropic-like plasma permeating the interstellar space}

\author{Filipe S. Ribeiro\orcid{0000-0003-4142-4304}$^a$}
\email{filipe.ribeiro@discente.ufma.br, filipe99ribeiro@hotmail.com}
	\author{Pedro D. S. Silva\orcid{0000-0001-6215-8186}$^b$}
	\email{pedro.dss@ufma.br, pdiego.10@hotmail.com}
		\author{Rodolfo Casana\orcid{0000-0003-1461-3038}$^{a,c}$}
	\email{rodolfocasana@gmail.com, rodolfo.casana@ufma.br}
	\author{Manoel M. Ferreira Jr.\orcid{0000-0002-4691-8090}$^{a,c}$}
	\email{manojr.ufma@gmail.com, manoel.messias@ufma.br}
		\affiliation{$^a$Programa de P\'{o}s-graduaç\~{a}o em F\'{i}sica, Universidade Federal do Maranh\~{a}o, Campus
		Universit\'{a}rio do Bacanga, S\~{a}o Lu\'is, Maranh\~ao 65080-805, Brazil}
		\affiliation{$^{b}$Coordena\c{c}\~ao do Curso de Ci\^encias Naturais - F\'isica, Universidade Federal do Maranh\~ao, Campus de Bacabal, Bacabal, Maranh\~ao, 65700-000, Brazil}
	\affiliation{$^c$Departamento de F\'{i}sica, Universidade Federal do Maranh\~{a}o, Campus
		Universit\'{a}rio do Bacanga, S\~{a}o Lu\'is, Maranh\~ao 65080-805, Brazil}

\begin{abstract}
In this work, we study the propagation of electromagnetic waves in a magnetized chiral plasma that pervades the interstellar space. The Maxwell equations, supplemented by bi-isotropic-like constitutive relations, are rewritten to describe a cold, uniform, and collisionless plasma model that yields new collective electromagnetic modes for distinct pairs of refractive indices associated with right- and left-handed circularly polarized waves. We have investigated the optical behavior through the rotatory power (RP) and dichroism coefficient, reporting that the finite chiral parameter induces double RP sign reversal, an exotic optical signature that takes place in chiral dielectrics and rotating plasmas. In the low-frequency regime, a modified propagating helicon with right-handed circular polarization is obtained.  Next, supposing that the interstellar medium behaves as a chiral bi-isotropic-like cold plasma, we employ Astrophysical data of radio pulsars to achieve upper limits on the magnetoelectric parameters magnitude.
In particular, by using dispersion measure and rotation measure data from five pulsars, we constrain the magnitude of the chiral parameter to the order of $10^{-16}$ and $10^{-22}$, respectively.

\end{abstract}
\pacs{11.30.Cp, 41.20.Jb, 41.90.+e, 42.25.Lc}
	\maketitle

\section{Introduction \label{themodel1}}

Chiral optically active media are characterized by the absence of inversion symmetry (parity), being described by parity-odd electromagnetic models \cite{Barron2, Hecht, Wagniere}. Optical activity makes left- and right-handed circularly polarized (LCP and RCP) waves propagate with distinct phase velocities, implying birefringence \cite{Fowles}, phenomenon that may arise from natural features of the medium or can be induced by external fields (e.g., Faraday effect \cite{Bennett, Porter, Shibata}). Birefringence is generally quantified in terms of the rotation angle per unit length or rotatory power (RP) \cite{Condon}, providing a useful way to probe nonusual electromagnetic responses \cite{Pesin, PedroPRD2020} and also to discuss underlying physics in context of particle physics and cosmology \cite{Petropavlova}. To account for such an effect in chiral systems and other optical-related phenomena, extensions of the standard isotropic description of the medium's electromagnetic response are often used \cite{Wang-Mirmoosa, Band, Mirmoosa-Mostafa,Martin-Ruiz}. For instance, in recent decades, bi-isotropic \cite{Sihvola,Sihvola_2} and bianisotropic electrodynamics \cite{Kong,Bianiso,Mahmood,Lorenci,Pedro3} have been considered to describe optically active systems, whose constitutive relations can be generally written as
\begin{subequations}
	\label{constitutive2}
	\begin{align}
	\mathbf{D}& =\hat{\epsilon}\, \mathbf{E}+\hat{\alpha}\,
	\mathbf{B},  \label{constitutive2a} \\
	\mathbf{H}& =\hat{\beta}\, \mathbf{E}+\hat{\mu}^{-1}\,\mathbf{B },
	\label{constitutive2b}
	\end{align}
\end{subequations}
where the constitutive tensors, {$\hat{\epsilon}=[\epsilon_{ij}]$, $\hat{\mu}^{-1}=[\mu^{-1}_{ij}]$, $\hat{\alpha}=[\alpha_{ij}]$ and $\hat{\beta}=[\beta_{ij}]$,}
represented by $3\times 3$ complex matrices, {describe electric permittivity, magnetic permeability, and magnetoelectric tensors, respectively. These quantities satisfy the following Hermiticity conditions:}
		\begin{align}
		\hat{\epsilon}=\hat{\epsilon}^{\dagger}, \quad
		\hat{\mu}^{-1}=\left(\hat{\mu}^{-1}\right)^{\dagger}, \quad
		\hat{\alpha}=-\hat{\beta}^{\dagger} \label{energycondition},
		\end{align}		
{which follow from energy conservation in continuous media} \cite{Kong,Kamenetskii}. The bi-isotropic relations involve {isotropic tensors}, ${\epsilon_{ij}}=\epsilon\delta_{ij}$, ${\alpha_{ij}}
	=\alpha\delta_{ij}$, ${\beta_{ij}}=\beta\delta_{ij}$, describing the most general linear, homogeneous, and isotropic materials with magnetoelectric coefficients. In this case, to assure energy conservation, the complex magnetoelectric coefficients fulfill
$\alpha=-\beta^{*}$.

Bi-isotropic constitutive relations are widely employed in modified electromagnetic field theories and condensed matter physics to describe optical properties of materials \cite{Aladadi, Jelinek,Chang}, the electromagnetic response of topological insulators \cite{Zou,Urrutia, Lakhtakia, Winder,Li,Li1,Tokura}, axion couplings \cite{Sekine,Nenno,Tobar,Kurumaji}, Casimir effect in chiral media \cite{Casimir,Schoger}, and applications in photonic systems \cite{Silveirinha}. Bi-anisotropic magnetoelectric parameters are generally encoded in the off-diagonal elements of the matrices $\alpha_{ij}$, $\beta_{ij}$ in relations (\ref{constitutive2a}) and (\ref{constitutive2b}), being used for investigating surface plasmon polaritons on bianisotropic substrates \cite{Darinskii}, group and energy velocities waves in heterostructures and surfaces \cite{Darinskii2}, wave propagation in time-dependent media with antisymmetric magnetoelectric coupling \cite{Lin}, Weyl semimetals \cite{Halterman, Zu}, ferromagnetic gyromagnetic materials \cite{ Krupka1}, discovery of electromagnons in perovskites associated with giant directional dichroism \cite{Takahashi}, and enhanced gyrotropic birefringence in a helimagnet \cite{Iguchi}. Anisotropic magnetoelectric coefficients are also important to describe magneto-optical effects of topological insulators \cite{Chang1,Tse} and graphene compounds \cite{Crasee}.

Optical activity also takes place in cold magnetized plasmas when the Faraday effect manifests for wave propagation parallel to the external magnetic field. Such a system is described by the conventional linear cold plasma theory \cite{Bittencourt,Boyd,Chabert}, whose permittivity tensor,
\begin{equation}
	\varepsilon_{ij}  (\omega)=\varepsilon_{0}%
	\begin{bmatrix}
		S & -iD & 0\\
		iD & S & 0\\
		0 & 0 & P
	\end{bmatrix}, \label{7.1}
\end{equation}
is obtained for a magnetic background, $\mathbf{B}=B_{0}\hat{z}$. Here, $\epsilon_{ijk}$ is the Levi-Civita symbol with $\epsilon_{123}=1$, $S=1 - {\omega_{p}^{2}}/{(\omega^{2}-\omega_{c}^{2})}$, $ D= { \omega_{c} \omega_{p}^{2}}/ ({\omega (\omega^{2} - \omega_{c}^{2})})$, and $P = 1-{\omega_{p}^{2}}/{\omega^{2}},$ in which $\omega_{p}^{2}=nq^{2}/(\varepsilon_{0}m)$ and $\omega_{c}=qB_{0}/m$ are the squared plasma and cyclotron frequencies for particles of mass $m$ and charge $q$, respectively. This system yields the well-known longitudinal plasma oscillation at $\omega=\omega_{p}$, and the transversal left- and right-handed circularly polarized modes, associated respectively with the refractive indices
\begin{equation}
	n_{\pm}= \sqrt{1- \omega_{p}^{2}/\omega\left(\omega\pm\omega_{c}\right)}. \label{nusual2}
\end{equation}

Plasma physics is closely related to two relevant astrophysical observables used for
 examining dispersion and polarization rotation of waves coming from distant sources: the dispersion measure (DM) and rotation measure (RM). For radio waves traveling a distance $d$ through the interstellar medium (ISM) described by a cold, ionized bi-isotropic chiral plasma, one defines the arrival time, $t=\int_{0}^{d}(v_{g})^{-1}ds$, where $s$ is the line of sight element and $v_{g}$ the group velocity \cite{Lyne}. For frequencies large compared to the plasma frequency, i.e, $\omega\gg\omega_{p}$, the ISM does not present absorption (lossless media). In this range, the {usual} plasma refractive indices\footnote{We consider an electron plasma, where the cyclotron frequency $\omega_{c}$ incorporates the electron charge $q = -e$.} (\ref{nusual2}) provide
\begin{equation}
	v_{g}^{-1}\approx \frac{1}{c}+\frac{\omega_{p}^{2}}{2  c\omega^{2}}\pm\frac{\omega_{c}\omega_{p}^{2}}{c\omega^{3}}, \label{groupvelocity}
\end{equation}
whose term in $\omega^{-3}$ is smaller than the second one by the factor $\omega_{c}/\omega \sim 10^{-7}$ (for an interstellar magnetic field of the order of a few $\mu \mathrm{G}$). Thus, it can be neglected, yielding the following arrival time
\begin{equation}
	t\approx \frac{d}{c}+\frac{e^{2}}{2c\epsilon_{0}m\omega^{2}}\text{DM}, \quad \text{DM}=\int^{d}_{0}n_{e}ds. \label{timedelay2}
\end{equation}
The dispersion measure (DM) is defined in terms of the electron number density, $n_{e}$, that represents the column of electrons along the path of integration.  The time delay is obtained by taking the difference between the transit time of two signals (one traveling at light speed $c$, another at $v_g$), that is,
\begin{equation}
	\tau = \frac{e^{2}}{2c\epsilon_{0}m\omega^{2}}\text{DM}, \label{timedelay3}
\end{equation}
marked by the well-known $\omega^{-2}$ behavior for electromagnetic signals.

The radio signals from pulsars also experience Faraday rotation due to the influence of the Galactic magnetic field. In this scenario, the differential phase rotation along the line of sight is written as
\begin{equation}
	\Delta\Psi=\int_{0}^{d}\left(  k_{R}-k_{L}\right)  ds=\frac{e^{3}\lambda^{2}}{4\pi^{2}c^{3}m^{2}\epsilon_{0}}\int_{0}^{d}n_{e}B_{\parallel} ds, \label{Faraday}
\end{equation}
where $B_{\parallel}$ is the magnetic field parallel to the line of sight\footnote{It is the relevant configuration to examine the large-scale structure of the Galactic field  \cite{Lyne,Simard-Normandin}, by combining estimates of $B_{\parallel}$ for several pulsars along different lines of sight.},  usually given in $\mu$G, while the distance $d$ is taken in pc. The wave numbers {appearing in \eqref{Faraday}} are associated with the indices (\ref{nusual2}) {through $k=\omega n$}, and, under the small density hypothesis, read
\begin{equation}
	k_{R,L}\approx \frac{\omega}{c}-\frac{\omega_{p}^{2}}{2c\omega}\pm\frac
	{\omega_{c}\omega_{p}^{2}}{2c\omega^{2}}.
\end{equation}
The third term, despite smaller than the second one, is the leading order term for the phase shift of \eqref{Faraday} and determines the Faraday effect under the small density condition. Thus, the differential phase rotation (\ref{Faraday}) yields the polarization rotation angle
\begin{equation}
	\Delta\phi \equiv\Delta\Psi/2 =\lambda^{2}\text{RM},
\end{equation}
where the rotation measure (RM) is defined as
\begin{equation}
		\text{RM}=\frac{e^{3}}{8\pi^{2}c^{3}m^{2}\epsilon_{0}}\int_{0}^{d}n_{e}B_{\parallel} ds,\label{Faraday2}
\end{equation}
given in $\mathrm{rad}/\mathrm{m}^{2}$. From the electromagnetic time delay, relevant information to estimate the Galactic electron distribution permeating the ISM \cite{Rybicki, Simard-Normandin,Lyne} are collected, in which the DM is a key parameter for studying dispersive ISM effects along the wave path \cite{Krishnakumar-1}. On the other hand, the RM is a widely employed tool for estimating the magnitude and direction of the Galactic magnetic fields, derived from analyzing the polarization of light emitted by pulsars \cite{Han,Ng,Sullivan}. Observations of radio pulsars by multiple telescopes have provided datasets that include several relevant quantities of these objects, such as DM and RM. The LOFAR collaboration \cite{LOFAR} has published data on DMs, flux densities, and calibrated total intensity profiles for a subset of pulsars observed by its high-band antennas (110–188 MHz). Additionally, LOFAR has significantly enhanced the precision of RM measurements for low-frequency pulsars \cite{LOFAR_RM}. The effects of a chiral plasma permeating the ISM, on the DM and RM of light coming from pulsars, were recently investigated in Ref.~\cite{Filipe2025}, yielding upper bounds on the chiral factors.

One way of engendering a chiral plasma is to consider the chiral magnetic effect (CME)~\cite{Fukushima, Kharzeev1C},  a phenomenon arising from an imbalance between right- and left-handed fermions which induces an electrical current proportional to the magnetic field, $\mathbf{J}=\sigma_B\mathbf{B}$, where $\sigma_{B}$ represents the magnetic conductivity. The emergence of such an effect leads to chiral plasma instabilities~\cite{Akamatsu,Wang, Duari, Carignano} and interesting astrophysical repercussions \cite{Kamada}. Indeed, the CME plays an important role in the amplification and evolution of magnetic fields in the early Universe \cite{Tashiro,Pavlovic}, in which the cosmic medium behaves as a chiral plasma. The chiral magnetic instability was further investigated in a protoneutron star during its early cooling phase (after the supernova core collapse), where it contributes to generate magnetic fields as large as $10^{14}$ G \cite{Sigl}. {The CME can also occur} in the magnetosphere of magnetars, triggering chiral plasma instability that can be associated with the emission of circularly polarized electromagnetic radiation in a wide window of frequencies, possibly affecting signals from pulsars and fast radio bursts features~\cite{Gorbar}. A recent investigation has shown that the chiral magnetic current acts analogously to a dynamo, amplifying magnetic fields by using the energy stored in the chiral chemical potential, providing toroidal and poloidal large-scale dipolar magnetic components of $10^{14}$ G \cite{Dehman}.  Furthermore, parity-violating effects may emerge from a chiral primordial gravitational-wave background \cite{GW1, GW2, GW3, GW4, GW5, GW6} that would induce chiral photons through photon-graviton interaction \cite{GW7}, contributing to the CMB polarization \cite{GW8}.

The chiral magnetic current can be effectively described within the
	framework of the \textit{CPT}-odd  Maxwell-Carroll-Field-Jackiw
	electrodynamics (MCFJ)~\cite{CFJ,Colladay, Qiu}, a Lorentz-violating extension of Maxwell theory whose Lagrangian\footnote{Here, $F_{\mu\nu}=\partial_{\mu}A_{\nu}-\partial_{\nu}A_{\mu}$ and  $G^{\mu\nu}=\frac{1}{2}\chi^{\mu\nu\alpha\beta}F_{\alpha\beta}$
		are the usual $U(1)$ vacuum and continuous matter field strength, respectively, where $\chi^{\mu\nu\alpha\beta}$ describes the medium constitutive tensor. }
	\begin{equation}
		\mathrm{{\mathcal{L}}}=-\frac{1}{4}G^{\mu \nu }F_{\mu \nu }+\frac{1}{4}%
		\epsilon ^{\mu \nu \alpha \beta }\left( k_{AF}\right) _{\mu }A_{\nu
		}F_{\alpha \beta },
	\end{equation}
	yields the inhomogeneous equations
	\begin{align}
		\nabla \cdot \mathbf{D}+\mathbf{k}_{AF}\cdot \mathbf{B}& =0,
		\label{Coulomb1} \\
		\nabla \times \mathbf{H}-\frac{\partial \mathbf{D}}{\partial t}& =-k_{AF}^{0}%
		\mathbf{B}+\mathbf{k}_{AF}\times \mathbf{E}.  \label{Amp1}
	\end{align}%
The modified Ampere's law incorporates anomalous current terms, where the factor $k_{AF}^{0}$ describes the magnetic conductivity and the vector $\mathbf{k}_{AF}$ represents the anomalous Hall conductivity. For the
	vacuum case, one writes
	\begin{align}
		\epsilon _{0}\nabla \cdot \mathbf{E}& =-\mathbf{k}_{AF}\cdot \mathbf{B},
		\label{GMCFJ} \\
		\frac{1}{\mu _{0}}\nabla \times \mathbf{B}-\epsilon _{0}\frac{\partial
			\mathbf{E}}{\partial t}& =-k_{AF}^{0}\mathbf{B}+\mathbf{k}_{AF}\times
		\mathbf{E}.  \label{AMCFJ}
\end{align}

\bigskip

The CFJ Lagrangian can also be obtained from the axion
	coupling,
	\begin{equation}
		\mathrm{{\mathcal{L}}}_{axion}=-\frac{1}{4}F^{\mu \nu }F_{\mu \nu }+\theta
		\left( \mathbf{r},t\right) \left( \mathbf{E}\cdot \mathbf{B}\right) ,
	\end{equation}%
	by means of an integration, which reveals the connection $\partial
	_{\mu }\theta =-\left( k_{AF}\right) _{\mu }$~\cite{Sekine}. The axion
	electrodynamics can also be addressed by introducing bi-isotropic
	constitutive relations~\cite{ChengGUO, Barredo, Barredo2},
	\begin{align}
		\mathbf{D}& =\epsilon_0 \mathbf{E}+\theta \mathbf{B},  \label{CR1A} \\
		\mathbf{H}& =\mu _{0}^{-1}\mathbf{B}-\theta \mathbf{E},  \label{CR1B}
	\end{align}%
	into the Maxwell's equations. In fact, by replacing the constitutive
	relation (\ref{CR1A}) in the Gauss's law, $\nabla \cdot \mathbf{D}=\rho ,$ one obtains
		\begin{equation}
			\epsilon_0 \nabla \cdot \mathbf{E}+(\nabla \theta )\cdot \mathbf{B}=\rho ,
	\end{equation}
	since $\nabla \cdot \mathbf{B=0}$. In the absence of
	sources, $\rho =0,$ the latter equation recovers Eq.~(\ref{GMCFJ}), with  $%
	\nabla \theta =\mathbf{k}_{AF}.$ At the same time, replacing the
	constitutive relations (\ref{CR1A}) and (\ref{CR1B}) in the usual Ampere's law, yields
	\begin{equation}
		\frac{1}{\mu _{0}}\nabla \times \mathbf{B}-\epsilon_0 \frac{\partial \mathbf{E}%
		}{\partial t}=\nabla \theta \times \mathbf{E}+\left(\frac{\partial \theta }{%
			\partial t}\right)\mathbf{B}+\mathbf{j},
\end{equation}
which is equivalent to Eq. (\ref{AMCFJ}) with  $\partial
	_{t}\theta =-\left( k_{AF}\right) _{0}$, $\nabla \theta =\mathbf{k}_{AF}$, and $\mathbf{j}=0$. These conditions suggest a possible equivalence between the MCFJ model and the bi-isotropic electrodynamics of Eqs.~(\ref{CR1A}) and (\ref{CR1B}). A mapping between the magnetoelectric parameter and the magnetic conductivity (or CFJ timelike factor) is given
	\begin{equation}
		\theta =i\left( k_{AF}\right) _{0}/\omega. \label{equival}
	\end{equation}
	The CFJ theory with $\left( k_{AF}\right) _{0}\neq 0$ provides an electromagnetic framework that describes plasmas with CME \cite{Filipe1,Filipe2}, phenomenon with astrophysical applications/effects, which can be connected with the bi-isotropic plasma addressed in the present work, given the equivalence stated in Eq.~(\ref{equival}).

As is well known, bi-isotropic electrodynamics constitutes an effective tool to describe real phenomena optical properties and wave propagation in chiral matter. On the other hand, in this work  we examine the possibility of the plasma being described by such extended constitutive relations. We develop such an investigation analyzing optical properties of a magnetized cold plasma ruled by bi-isotropic-like constitutive relations (\ref{constitutiveRELATIONS2}). In Sec.~\ref{Propag.TL}, we determine the modified refractive indices and analyze the propagation and absorption zones.  In Sec.~\ref{birefringence}, we discuss optical effects of the chiral plasma, such as birefringence and dichroism, supposing a finite magnitude chiral factor. In Sec.~\ref{Sec_Astrophysical_Aplication}, astrophysical data of DM and RM of five pulsars are employed to constrain the magnitude of the chiral parameter, assuming that radio waves travel in an interstellar bi-isotropic-like medium (ISM). Finally, we summarize the results and perspectives in Sec.~\ref{conclusion}.

\section{Wave propagation in bi-isotropic-like chiral plasma}
\label{Propag.TL}

Chiral cold plasmas {are another type of system exhibiting optical activity and have been investigated in at least two distinct scenarios: i) in the} axionlike MCFJ electrodynamics \cite{CFJ,Colladay}, considering the influence of the chiral magnetic current and anomalous Hall current on the propagating modes and optical features \cite{Filipe1,Filipe2,Belich}; ii) in bi-isotropic-like Maxwell electrodynamics \cite{Gao,Guo,Silva}, where negative refraction emerges as a consequence of the chiral parameter. In this latter case, the electric permittivity is given the tensor of \eqref{7.1} and the constitutive relations are
\begin{subequations}
	\label{constitutiveRELATIONS2}
	\begin{align}
		\mathbf{D}^{i}  & =\epsilon_{ij}(\omega)\mathbf{E}^{j}+i\xi_{c}\mathbf{B}^{i},\label{1BB}\\
		\mathbf{H}^{i}  &  =i\xi_{c}\mathbf{E}^{i}+\mu_0^{-1}\mathbf{B}^{i}, \label{2BB}%
	\end{align}
\end{subequations}
which hold under the condition (\ref{energycondition}). Notice that replacing \eqref{constitutiveRELATIONS2} in the usual Maxwell equations, one finds
\begin{align}
\frac{1}{\mu_0} (\nabla \times {\bf{B}})^{i} - \epsilon_{ij}(\omega) \partial_{t} {\bf{E}}^{j} &= 2 i \xi_{c} \partial_{t} {\bf{B}}^{i} , \label{extra-connection-1}
\end{align}
which allows us to establish a connection between $\xi_{c}$ and the CFJ timelike factor as
\begin{align}
2\xi_{c} \omega = - k_{AF}^{0} .  \label{extra-connection-2}
\end{align}
A similar correspondence has been pointed in Ref.~\cite{Dossow-Urrutia}. Considering \eqref{equival} and \eqref{extra-connection-2}, one can interpret that the CFJ timelike factor can effectively describe a dispersive chiral electromagnetic response of bi-isotropic systems.

In the present work, we examine a chiral plasma described by the bi-isotropic-like constitutive relations of \eqref{1BB} and \eqref{2BB}, that replaced in the Maxwell equations, yields the modified wave equation
\begin{equation}
	\left[n^{2}\delta_{ik}-n^{i}n^{k}-2i\tilde{\xi}_{c}\epsilon_{ijk}n^{j}-
\varepsilon_{ik}\left(\omega\right)/\varepsilon_{0}\right]  E^{k}=0, \label{EWE_bi_isotropic}
\end{equation}
where $\tilde{\xi}_{c}=\xi_{c}/(c\varepsilon_{0})$ is the redefined chiral parameter, and $\varepsilon_{ik}\left(\omega\right)$ is the plasma permittivity tensor, given in \eqref{7.1}.

In this section, we derive the collective electromagnetic modes for a cold magnetized bi-isotropic plasma, whose wave equation (\ref{EWE_bi_isotropic}) provides its dispersive properties. Such an equation can be read in a matrix form,
\begin{widetext}
	\begin{equation}
	\begin{bmatrix}
	n^{2}-n_{x}^{2}-S & -n_{x}n_{y}+2i\tilde{\xi}_{c}n_{z} +iD&
	-n_{x}n_{z}-2i\tilde{\xi}_{c}n_{y}\\
	-n_{x}n_{y}-2i\tilde{\xi}_{c}n_{z}-iD & n^{2}-n_{y}^{2}-S & -n_{y}n_{z}+2i\tilde{\xi}_{c}n_{x}\\
	-n_{x}n_{z}+2i\tilde{\xi}_{c}n_{y} & -n_{y}n_{z}-2i\tilde{\xi}_{c}n_{x} & n^{2}-n_{z}^{2} -P
	\end{bmatrix}
	\begin{bmatrix}
	E_{x}\\
	E_{y}\\
	E_{z}%
	\end{bmatrix}
	=0.\label{Matrix_bi_isotropic}
	\end{equation}
\end{widetext}

For simplicity, let us consider a coordinate system where the refractive index is parallel to the external magnetic field, $\mathbf{n}=n\hat{z}$, setting up the so-called Faraday configuration, for which \eqref{Matrix_bi_isotropic} reads
\begin{equation}
\begin{bmatrix}
n^{2}-S & 2i\tilde{\xi}_{c}n +iD&
0\\
-2i\tilde{\xi}_{c}n-iD & n^{2}-S & 0\\
0 & 0 & -P
\end{bmatrix}
\begin{bmatrix}
 E_{x}\\
 E_{y}\\
E_{z}%
\end{bmatrix}
=0, \label{Matrix_bi_isotropic2}
\end{equation}
whose nontrivial solutions arise from the condition that the determinant of the coefficient matrix vanishes, yielding
	\begin{equation}
	P \left[
	\left(  n^{2}-S\right)  ^{2}-\left(2\tilde{\xi}_{c}n+D\right)^{2}\right]=0.
	\label{DR1A}
	\end{equation}
The above dispersion relation describes the dispersive properties of the system and determines its allowed propagating modes. For longitudinal wave configuration, $\mathbf{E}=(0,0, E_{z})$,  the standard non propagating mode stems from $P=0$ and occurs at the plasma frequency, $\omega=\omega_{p}$, also inducing a longitudinal collective mode of electrons, called Langmuir oscillations.

For transverse waves, $\mathbf{n}\perp\mathbf{E}$ and $\mathbf{E}=(\delta E_{x},\delta E_{y},0)$, Eq.~(\ref{DR1A}) provides
\begin{equation}
n^{2}\pm\left(2\tilde{\xi}_{c}n+D\right)-S  =0,
\end{equation}
with $D$ and $S$ defined after Eq.~(\ref{7.1}), and yields the refractive indices
\begin{align}
	n_{\text{R}_{\pm}} &= - \tilde{\xi}_{c} \pm \sqrt{ 1 +\tilde{\xi}_{c}^{2} - \frac{\omega_{p}^{2}}{\omega\left(\omega-\omega_{c}\right)}} , \label{n-R-M-indices-1_bi_isotropic} \\
	n_{\text{L}_{\pm}} &= \tilde{\xi}_{c} \pm \sqrt{ 1 + \tilde{\xi}_{c}^{2} - \frac{\omega_{p}^{2}}{\omega\left(\omega+\omega_{c}\right)}}, \label{n-L-E-indices-1_bi_isotropic}
	\end{align}
associated with circularly polarized modes. One obtains a LCP mode, associated with the indices $n_{L_{+}}$ and $n_{L_{-}}$,
\begin{equation}
n_{L_{+}}, n_{L_{-}}  \   \mapsto  \ \mathbf{{E}}_{LCP}=\frac{i}{\sqrt{2}}%
\begin{bmatrix}
1 \\
i\\
0
\end{bmatrix}, \label{ELCP2}
\end{equation}
and a RCP mode, connected with the indices $n_{R_{+}}$ and $n_{R_{-}}$,
\begin{equation}
n_{R_{+}}, n_{R_{-}} \   \mapsto  \ \mathbf{{E}}_{RCP}=\frac{i}{\sqrt{2}}%
\begin{bmatrix}
1 \\
-i\\
0
\end{bmatrix}.\label{ERCP2}
\end{equation}

\begin{figure}[h]
	\centering
	\includegraphics[scale=0.7]{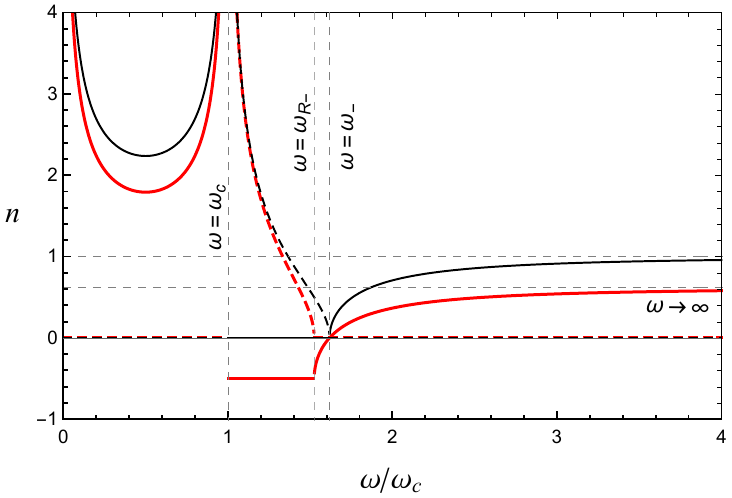}  \caption{Frequency dependence of the index of refraction $n_{R_{+}}$ for $\omega_{p}=\omega_{c}$. The solid (dashed) red line corresponds to the real (imaginary) piece of $n_{R_{+}}$. The same pattern holds for index ${n}_{-}$ of \eqref{nusual2}, shown here for comparison. Here, $\tilde{\xi_{c}}=0.8$ and $\omega_{p}=1$~$\mathrm{rad}$~$s^{-1}$.}
	\label{NR+_bi-isotropic}%
\end{figure}

\begin{figure}[h]
	\centering
	\includegraphics[scale=0.7]{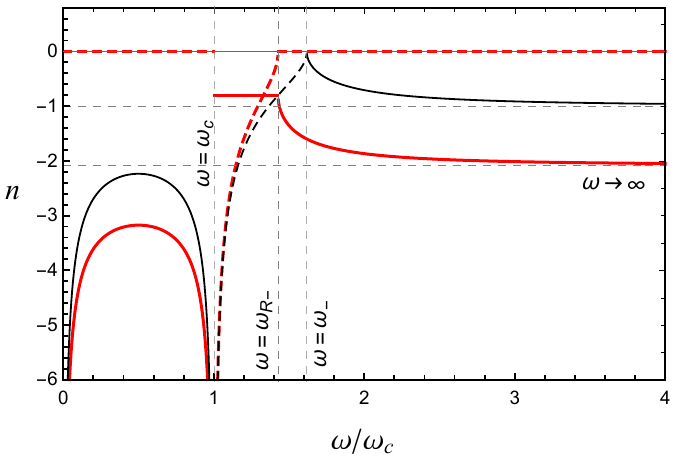}  \caption{{Frequency dependence behavior of $n_{R_{-}}$} for $\omega_{p}=\omega_{c}$. The dashed red (black) line corresponds to the imaginary piece of $n_{R_{-}}$ ($-{n}_{-}$), while the solid red (black) line represents the real piece of $n_{R_{-}}$ ($-{n}_{-}$). The index ${n}_-$ is given in \eqref{nusual2}. Here, $\tilde{\xi}_{c}=0.8$ and $\omega_{p}=1$~$\mathrm{rad}$~$s^{-1}$.}
	\label{NR-_bi-isotropic}
\end{figure}

\begin{figure}[h]
	\centering
	\includegraphics[scale=0.7]{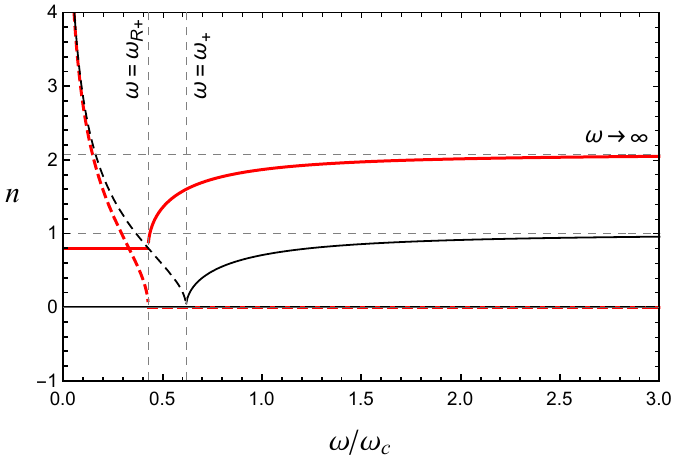}  \caption{Index of refraction
			$n_{L_{+}}$ for $\omega_{p}=\omega_{c}$. The dashed red (black) line corresponds to the imaginary piece of $n_{L_{+}}$ ($\tilde{n}_{+}$), while the solid red (black) line represents the real piece of $n_{L_{+}}$ ($\tilde{n}_{+}$), where the index $\tilde{n}_+$ is given in \eqref{nusual2}. Here, $\tilde{\xi}_{c}=0.8$ and $\omega_{p}=1$~$\mathrm{rad}$~$s^{-1}$.}
	\label{NL+_bi-isotropic}%
\end{figure}

\begin{figure}[h]
	\centering
	\includegraphics[scale=0.7]{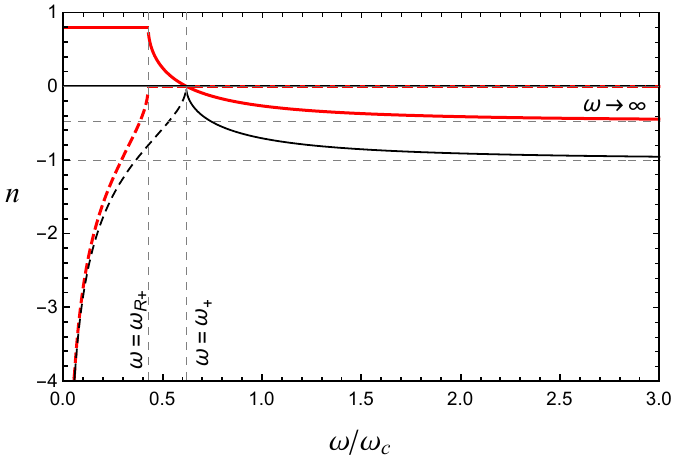}  \caption{Index of refraction
			$n_{L_{-}}$ for $\omega_{p}=\omega_{c}$. The dashed red (black) line corresponds to the imaginary piece of $n_{L_{-}}$ ($\tilde{n}_{+}$), while the solid red (black) line represents the real piece of $n_{L_{-}}$ ($\tilde{n}_{+}$). The index $\tilde{n}_+$ is given in \eqref{nusual2}. Here, $\tilde{\xi_{c}}=0.8$ and $\omega_{p}=1$~$\mathrm{rad}$~$s^{-1}$.}
	\label{NL-_bi-isotropic}%
\end{figure}

Standard cutoff frequencies, namely
\begin{equation}
\label{standard-cutoff-frequency-1}
	\omega_{\pm}=  \frac{1}{2}\left(\sqrt{4\omega_{p}^{2}  +\omega_{c}^{2}} \mp \omega_{c}\right),
\end{equation}
occurs in the refractive indices $n_{R_{+}}$ and $n_{L_{-}}$, given by $\omega_{+}$ and $\omega_{-}$, respectively. On the other hand, the indices  $n_{R_{-}}$ and $n_{L_{+}}$ have no real roots, {being completely negative and positive for all frequency values, respectively}. It is important to notice that the indices (\ref{n-R-M-indices-1_bi_isotropic}) and (\ref{n-L-E-indices-1_bi_isotropic}) become complex\footnote{{As is well known}, purely imaginary or complex (presenting both real and imaginary pieces) refractive indices indicate absorption of electromagnetic waves, {whereas} real indices are associated with {attenuation-free} propagating waves.} in the range in which the radicands,
 \begin{equation}
 R_{\pm}\left(\omega\right)= 1 +\tilde{\xi}_{c}^{2} - \frac{\omega_{p}^{2}}{\omega\left(\omega\pm\omega_{c}\right)},\label{Radicand_bi}
 \end{equation}
are negative. These regions are limited by one of the roots,
 \begin{equation}
 \label{roots-frequency-1}
\omega_{R\pm}=  \mp \omega_{c}/2+\sqrt{\omega_{p}^{2}/ (
	1+\tilde{\xi}_{c}^{2} )  +\omega_{c}^{2}/4} ,
\end{equation}
as shown in the plots of Figs.~\ref{NR+_bi-isotropic},~\ref{NR-_bi-isotropic},~\ref{NL+_bi-isotropic},~\ref{NL-_bi-isotropic}. {Note that $\omega_{R\pm}$ is related to $n_{R\pm}$ and $n_{L\pm}$, respectively.}

Now, we examine the main impacts of the chiral parameter $\tilde{\xi_{c}}$ on the refractive indices supposing its magnitude is comparable with the optical observables of the system {(in order to magnify its action)}. These features are pointed out below:
\begin{itemize}
	\item The refractive index $n_{R_{+}}$, illustrated in Fig.~\ref{NR+_bi-isotropic}, presents a free propagation (without absorption) zone for $0<\omega<\omega_{c}$, similarly to the usual case (solid black line). For $ \omega_{c}<\omega<\omega_{R-}$, the index is complex, $\mathrm{Im}[n_{R_{+}}]\ne 0$, and an absorption zone occurs (where the real part of the index is negative). As $\omega_{R-}<\omega_{-}$, one has a shortened absorption zone compared to the usual case, see Fig.~\ref{NR+_bi-isotropic}. In addition, a free propagating negative refraction zone {(nonexistent in the standard case)} emerges for $\omega_{R-}<\omega<\omega_{-}$.

	\item The negative refractive index $n_{R_{-}}$ is depicted in Fig.~\ref{NR-_bi-isotropic}, and shows a propagation zone for $0<\omega<\omega_{c}$ and a shortened absorption zone for the interval $\omega_{c}<\omega<\omega_{R-}$, {with real part of $n_{R-}$ being negative. The purely metamaterial-like} propagation zone holds for $\omega_{R-}<\omega<\omega_{-}$.
	
	 \item For the refractive index $n_{L_{+}}$, exhibited in Fig. \ref{NL+_bi-isotropic}, a partial absorption zone occurs for $0<\omega<\omega_{R+}$,  while for $\omega>\omega_{R+}$ {attenuation-free propagation takes place.}

	 \item For the refractive index $n_{L_{-}}$, illustrated in Fig. \ref{NL-_bi-isotropic}, it occurs partial absorption for $0<\omega<\omega_{R+}$, followed by an unusual anomalous refraction ($dn/d\omega <0$) propagation zone for the interval $\omega_{R+}<\omega<\omega_{+}$, where  $\mathrm{Re}[n_{L_{-}}]> 0$. For $\omega>\omega_{+}$, the index is purely real and negative, representing a negative refraction propagating zone.

\end{itemize}
In the high-frequency limit, the bi-isotropic plasma indices (\ref{n-R-M-indices-1_bi_isotropic}) and (\ref{n-L-E-indices-1_bi_isotropic}) tend to the asymptotical values  $n_{R_{\pm}}\to-\tilde{\xi_{c}}\pm\sqrt{1+\tilde{\xi}_{c}^{2}}$ and $n_{L_{\pm}}\to\tilde{\xi_{c}}\pm\sqrt{1+\tilde{\xi}_{c}^{2}}$, {differing} from the vacuum values, $n_{\pm}\to 1$, {which holds} in the conventional plasma scenario.

\subsection{\label{section-helicons}Low-frequency modes}

Helicons are RCP modes that propagate at very low frequencies,
\begin{align}
	\omega\ll \omega_{p}, \quad \omega_{c}\ll\omega_{p}, \quad  \omega\ll \omega_{c}, \label{helicon-frequency-regime_bi}
\end{align}
and along the {magnetic field axis \cite{Bittencourt, Chabert}}, being associated with the following refractive index:
\begin{align}
	n_{-} &= \omega_{p} \sqrt{\frac{1}{\omega \omega_{c}}}, \label{helicons-12_bi}
\end{align}
For the bi-isotropic magnetized chiral plasma described by the indices (\ref{n-R-M-indices-1_bi_isotropic}) and (\ref{n-L-E-indices-1_bi_isotropic}), the low frequency regime yields
\begin{align}
	\bar{n}_{R_{\pm}}&= \pm\omega_{p} \sqrt{\frac{1}{\omega \omega_{c}}}-\tilde{\xi_{c}}, \label{helicons-15_bi} \\
	\bar{n}_{L_{\pm}} &= \pm i\omega_{p} \sqrt{\frac{1}{\omega \omega_{c}}}+\tilde{\xi_{c}}, \label{helicons-16_bi}
	\end{align}
where we have used the ``bar'' notation to indicate the {helicon quantities}. The indices (\ref{helicons-15_bi}) and (\ref{helicons-16_bi}) {contain a linear and nondispersive chiral contribution.} The indices $\bar{n}_{R_{\pm}}$ are real, while $\bar{n}_{L_{\pm}}$ are complex, {in accordance  with the behavior observed in the low-frequency range for the refractive indices in Figs. \ref{NR+_bi-isotropic}--\ref{NL-_bi-isotropic}.} Thus, the RCP mode propagates, whereas the LCP mode is absorbed in the low-energy regime. {Furthermore, the index $\bar{n}_{R_{+}}$ may exhibit negative refraction zone when $\omega > \hat{\omega}$, with $\hat{\omega}=\omega_{p}^{2}/ (\tilde{\xi}_{c}^{2}\omega_{c}) $}. Thus, the conventional refraction ($\bar{n}_{R_{+}}>0$) for the RCP helicon is ensured only for $\omega<\hat{\omega}$.

{\section{\label{birefringence}Birefringence effects}}

{Circular birefringence} is characterized by the rotation of the wave polarization vector, {arising from the difference in the} phase velocities between RCP and LCP modes. In the present case, {this rotation effect is caused by the distinct phase velocities,  $v_{R_{\pm}}=1/n_{R_{\pm}}$ and $v_{L_{\pm}}=1/n_{L_{\pm}}$, associated with the indices (\ref{n-R-M-indices-1_bi_isotropic}) and (\ref{n-L-E-indices-1_bi_isotropic}), respectively.} In general, the birefringence is evaluated in terms of the rotatory power,
\begin{equation}
	\delta=-\frac{\omega}{2}\left(\mathrm{Re}\left[n_{LCP}\right]-
\mathrm{Re}\left[n_{RCP}\right]\right),
\end{equation}
where $n_{LCP}$ and $n_{RCP}$ are the refractive indices {for the LCP and RCP modes, respectively.} The rotatory power (RP) is usually employed in optical characterization of birefringent matter as organic compounds \cite{Xing-Liu}, graphene phenomena at terahertz band \cite{Poumirol}, gas of fast-spinning molecules \cite{Tutunnikov}, chiral metamaterials \cite{Woo, Zhang, Mun}, chiral semimetals \cite{Pesin, Dey-Nandy}, and in the determination of the rotation direction of pulsars \cite{Gueroult2,Gueroult}.

\subsection{Rotatory power \label{secRP}}

\begin{figure}[t]
	\centering
	\includegraphics[scale=0.69]{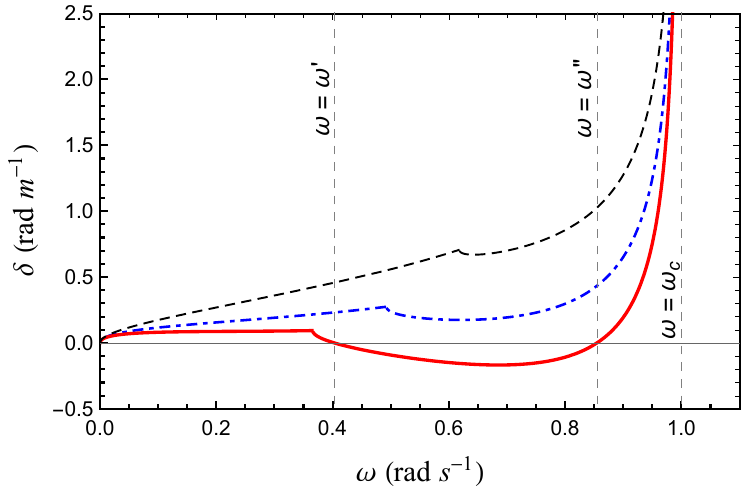}
	\caption{The rotatory power  (\ref{RP_RL+_bi}) for the interval $0<\omega<\omega_{c}$. It is illustrated in dot-dashed blue line ($\tilde{\xi_{c}}=0.6$) and solid red line ($\tilde{\xi_{c}}=1.0$). The standard plasma RP ($\tilde{\xi_{c}}=0$) is represented by the dashed black line. Here, we have used $\omega_{p}=1$~$\mathrm{rad}$~$s^{-1}$.}
	\label{Plot_RP2_bi}
\end{figure}

\begin{figure}[t]
	\centering
	\includegraphics[scale=0.69]{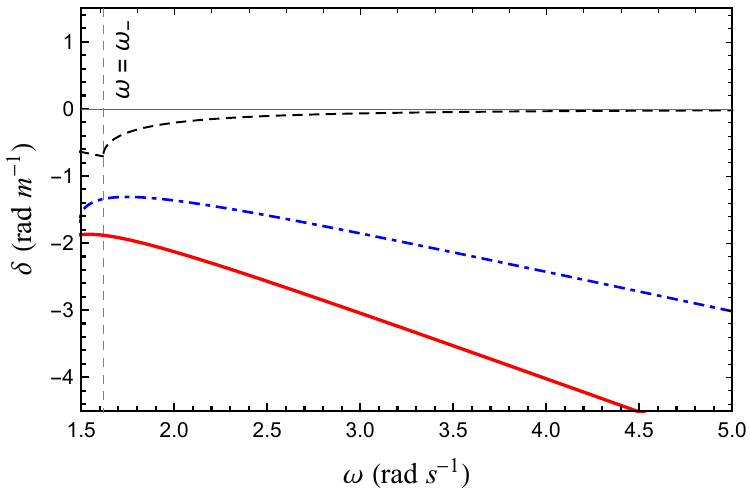}
	\caption{The rotatory power (\ref{RP_RL+_bi}) for the range $\omega>\omega_{-}$. It is represented by the dot-dashed blue line ($\tilde{\xi_{c}}=0.6$) and solid red line ($\tilde{\xi_{c}}=1.0$). The standard plasma RP  ($\tilde{\xi_{c}}=0$) is illustrated in a dashed black line. Here, we have used $\omega_{p}=1$~$\mathrm{rad}$~$s^{-1}$.}
	\label{Plot_RP1_bi}
\end{figure}

Considering the bi-isotropic chiral plasma is described by the indices $n_{L_{\pm}}$ and $n_{R_{\pm}}$, associated with the LCP and RCP waves, one can select waves that propagate in the same direction $+\hat{z}$, with $n>0$, to determine the RP. Thus, for this analysis, we will not consider $n_{R_{-}}$, which is always negative. Initially, we consider the indices $n_{L_{+}}$ and $n_{R_{+}}$, for which the RP is
	\begin{equation}
	\delta_{LR}^{++}=-\frac{\omega}{2} \left(2\tilde{\xi_{c}}+\mathrm{Re}\left[\sqrt{R_{+}\left(\omega\right)}\right]
-\mathrm{Re}\left[\sqrt{R_{-}\left(\omega\right)}\right]\right), \label{RP_RL+_bi}
	\end{equation}
where $R_{\pm}\left(\omega\right)$ are the radicands given in (\ref{Radicand_bi}). The behavior of the RP above is illustrated in Figs.~\ref{Plot_RP2_bi} and \ref{Plot_RP1_bi}, for $0<\omega<\omega_{c}$ and $\omega>\omega_{-}$, respectively\footnote{For the intermediary interval, $\omega_{c}<\omega<\omega_{-}$, the RP (\ref{RP_RL+_bi}) is not defined, since RCP and LCP waves are associated to opposite phase velocities (determined by the signals of the refractive indices), with $n_{R_{+}}<0$ and $n_{L_{+}}>0$ (see Figs. \ref{NR+_bi-isotropic} and \ref{NL+_bi-isotropic}).}, considering three values for the chiral parameter: $\tilde{\xi}_{c}=0$ (dashed black line), $\tilde{\xi}_{c}=0.8$ (dot dashed blue line), and $\tilde{\xi}_{c}=1.0$ (solid red line). For $0<\omega<\omega_{c}$, we note that the chiral parameter induces a double sign reversal, as occurs in the case in red line (see Fig.~\ref{Plot_RP2_bi}), with inversions occurring at $\omega^{\prime}$ and $\omega^{\prime\prime}$.  For $\omega_{c}<\omega<\omega_{-}$, the RP (\ref{RP_RL+_bi}) is always negative and decreases with the frequency for $\omega>\omega_{-}$ {[see Fig.~\ref{Plot_RP1_bi}]}. In the high-frequency regime, $\omega\gg\left(\omega_{c},\omega_{p}\right)$, the RP behaves as
\begin{equation}
\delta_{LR}^{++}\approx- \tilde{\xi_{c}}\omega-\frac{\omega_{c}\omega_{p}^{2}}{2\sqrt{1+\tilde{\xi_{c}}^{2}}\omega^{2}}.
\label{RP_high_frequency}
\end{equation}
exhibiting an unusual linear decay with frequency due to the chiral contribution in the first term.

\begin{figure}[t]
	\centering
	\includegraphics[scale=0.69]{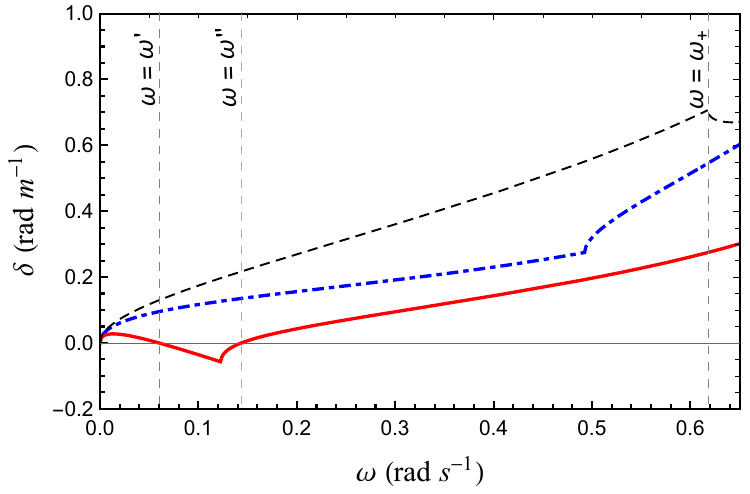}
	\caption{The rotatory power (\ref{RP_RL-+}) for the interval $0<\omega<\omega_{+}$. It is represented by the dot-dashed blue line ($\tilde{\xi_{c}}=0.6$) and solid red line ($\tilde{\xi_{c}}=2.5$), while the standard plasma RP ($\tilde{\xi_{c}}=0$) is illustrated in dashed black line. Here, we have used $\omega_{p}=1$~$\mathrm{rad}$~$s^{-1}$.}
	\label{Plot_RP3_bi}
\end{figure}

Considering the refractive indices $n_{L_{-}}$ and $n_{R_{+}}$, {which are both positive in the frequency range $0<\omega<\omega_{+}$},  the RP is given by
	\begin{align}
\delta_{LR}^{-+}&=-\frac{\omega}{2} \left(2\tilde{\xi_{c}}-\mathrm{Re}\left[\sqrt{R_{+}\left(\omega\right)}\right]-
\mathrm{Re}\left[\sqrt{R_{-}\left(\omega\right)}\right]\right), \label{RP_RL-+}
\end{align}	
{whose behavior} is depicted in Fig.~\ref{Plot_RP3_bi} for  $\tilde{\xi_{c}}=0$ (dashed black line), $\tilde{\xi_{c}}=0.8$ (dot dashed blue line), and $\tilde{\xi_{c}}=2.5$ (solid red line). {Notice that increasing the magnitude of the chiral parameter can also induce a double inversion at $\omega^{\prime}$ and $\omega^{\prime\prime}$, as illustrated in the red line of Fig.~\ref{Plot_RP3_bi}}. Therefore, the RP reversion appears in the bi-isotropic cold plasma for the two coefficients defined in this section, given in (\ref{RP_RL+_bi}) and (\ref{RP_RL-+}).

Rotatory power reversion is not a usual behavior in standard cold plasma theory. However, it has been observed in magnetized chiral cold plasmas ruled by the MCFJ electrodynamics \cite{Filipe1,Filipe2} and rotating plasmas rotating plasmas \cite{Gueroult2,Gueroult}. RP reversion {has also been} reported in graphene systems \cite{Poumirol} and bi-isotropic dielectrics supporting chiral magnetic current \cite{PedroPRB}. Recently, double rotatory power reversal has also been reported in bi-isotropic materials under the anomalous Hall effect (AHE), in the context of the axion electrodynamics \cite{Alex}.

\subsection{Dichroism coefficients \label{secDC} }

Circular dichroism occurs when circularly polarized waves experience differential absorption in media {whose refractive indices are complex. This effect is evaluated through} the coefficient
\begin{equation}
\delta_{d} =-\frac{\omega }{2}\left( \mathrm{Im}[n_{LCP}]-\mathrm{Im}[n_{RCP}]\right),  \label{dicro_eqtl}
\end{equation}
which takes into account the imaginary parts of the refractive indices associated with the LCP and RCP waves.

\begin{figure}
	\centering
	\includegraphics[scale=0.69]{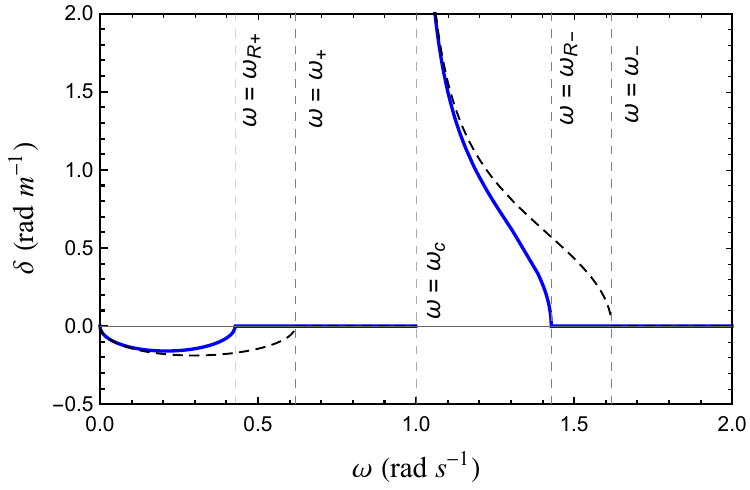}
	\caption{Dichroism coefficient of \eqref{dicro_bi_1}. The black dashed line represents the dichroism coefficient of a usual cold plasma.  Here, we have used $\tilde{\xi_{c}}=0.8$ and $\omega_{p}=1$~$\mathrm{rad}$~$s^{-1}$.}
	\label{Plot_DC1_bi}
\end{figure}

As discussed in Sec.~\ref{Propag.TL}, the refractive indices $n_{L_{\pm}}$ and $n_{R_{\pm}}$ become complex for $\omega<\omega_{R+}$ and $\omega_{c}<\omega<\omega_{R-}$, respectively, where absorption takes place. Taking into account the indices $n_{L_{+}}$ and $n_{R_{+}}$, the dichroism coefficient reads
\begin{equation}
\delta_{dLR}^{++}=\begin{cases}
\text{$ -\frac{\omega}{2} \sqrt{R_{+}\left(\omega\right)}$}, &\quad\text{for $0<\omega<\omega_{R+}$},\\
\text{$0$}, &\quad\text{for $\omega_{R+}<\omega<\omega_{c}$},\\
\text{$ +\frac{\omega}{2} \sqrt{R_{-}\left(\omega\right)}$}, &\quad\text{for $\omega_{c}<\omega<\omega_{R-}$},\\
\text{$0$}, &\quad\text{for $\omega>\omega_{R-}$}.\\
\end{cases}     \label{dicro_bi_1}
\end{equation}
whose behavior is exhibited in Fig.~\ref{Plot_DC1_bi}. The regions where the dichroism coefficient is non-null are shortened in comparison to the usual magnetized plasma, in accordance with the absorption bands of the associated refractive indices, as seen in Fig.~\ref{NR+_bi-isotropic} and Fig.~\ref{NL+_bi-isotropic}. A similar behavior {has been} obtained for a chiral plasma ruled by Maxwell-Carroll-Field-Jackiw theory, where two regions for non-null circular dichroism coefficient have appeared under specific conditions, as discussed in Refs.~\cite{Filipe1,Filipe2}.

Considering the indices  $n_{L_{-}}$ and $n_{R_{+}}$, the dichroism coefficient is
\begin{equation}
\delta_{dLR}^{-+}=\begin{cases}
\text{$ +\frac{\omega}{2} \sqrt{R_{+}\left(\omega\right)}$}, &\quad\text{for $0<\omega<\omega_{R+}$},\\
\text{$0$}, &\quad\text{for $\omega_{R+}<\omega<\omega_{c}$},\\
\text{$ +\frac{\omega}{2} \sqrt{R_{-}\left(\omega\right)}$}, &\quad\text{for $\omega_{c}<\omega<\omega_{R-}$},\\
\text{$0$}, &\quad\text{for $\omega>\omega_{R-}$},\\
\end{cases}     \label{dicro_bi_2}
\end{equation}
{is non-null in the same intervals of the first case and has the same magnitude as $\delta_{dLR}^{+,+}$, except for $0<\omega<\omega_{R+}$, where $\delta_{dLR}^{-,+}$ assumes opposite values, that is, $\delta_{dLR}^{-,+}= -|\delta_{dLR}^{+,+}|$.}

\section{Astrophysical constraints \label{Sec_Astrophysical_Aplication}}

In the latter section, we have examined the repercussions of the chiral parameter on the optical properties of cold plasma, purposefully supposing its magnitude is significant, as is commonly done in some bi-isotropic continuous media. In this section, however, our procedure is to estimate real upper limits for the chiral parameter magnitude, $\tilde{\xi}$. Thus, using well-established astrophysical DM and RM data from radio pulsar signals, we achieve such constraints under the procedure adopted in Ref.~\cite{Filipe2025}.

\subsection{Dispersion measure and rotation measure in chiral interstellar medium}

Considering the refractive indices $n_{R+}$ and $n_{L+}$, given in (\ref{n-R-M-indices-1_bi_isotropic}) and (\ref{n-L-E-indices-1_bi_isotropic}), the associated group velocities are\footnote{Here, we considered $q=-e$, the electron charge present in cyclotron frequency $\omega_{c}$.}
\begin{equation}
	\left(v_{g}\right)_{L,R}^{-1}\approx \frac{1}{c}\pm\frac{\tilde{\xi_{c}}}{c}+\frac{\omega_{p}^{2}}{2  \omega^{2}}, \label{groupvelocity_chiral}
\end{equation}
where {we have considered $\omega \gg  \omega_{p}$}, keeping the linear contribution in $\tilde{\xi}_{c}$. The time delay reads
\begin{equation}
	\tau= {\frac{e^{2}}{2c\varepsilon_{0}m\omega^{2}} } \left(  \text{DM}%
	\pm\text{DM}_{chiral}\right),\label{timedelay_Chiral}
	\end{equation}
and the $(\pm)$ signals are associated with LCP and RCP waves, respectively. We note a dispersive chiral contribution to the usual DM, proportional to the square of the frequency
	\begin{equation}
	\text{DM}_{chiral}=2\epsilon_{0}m\omega^{2}\tilde{\xi}_{c}d/e^{2}. \label{chiral_DM}
	\end{equation}
The wave vector associated to the refractive indices $n_{R+}$ and $n_{L+}$ are
	\begin{equation}
	k_{R,L}  = {\mp\omega\frac{\tilde{\xi}_{c}}{c} } +\frac{\omega}{c}\sqrt{1-\frac{\omega_{p}^{2}}{\omega\left(\omega\pm\omega_{c}\right)}
+\tilde{\xi}_{c}^{2}},\label{n_chiral}
	\end{equation}
which, in accordance with Eqs.~(\ref{Faraday}) and (\ref{Faraday2}), yields the Faraday rotation
\begin{equation}
	\Delta\phi  =\lambda^{2}\left(\text{RM}-\text{RM}_{chiral}\right),\label{Faraday_chiral}
	\end{equation}
with the wavelength-dependent chiral contribution,
\begin{equation}
\text{RM}_{chiral}=2\pi \tilde{\xi}_{c}d/\lambda^{3}.\label{RM_chiral}
\end{equation}

\subsection{Constraints using LOFAR dataset}

Observational dispersion measure values for several pulsars are available in the LOFAR census dataset \cite{LOFAR},  from which, for our estimates, we have selected five, namely, B1919+21, B1944+17,  B1929+10, B2016+28, and B2020+28. Uncertainties associated with each measure, here denoted by  $\epsilon_{\text{DM}}$, are also available in the catalog and will be assigned to the chiral parameter in our present proposal. Although it may consist of a nonrigorous approach, it can be employed to provide a first constraining estimate, justified by the assumption of a small chiral parameter in the bi-isotropic interstellar plasma. Therefore, for each measurement, we can constrain the magnitude of the chiral correction (\ref{chiral_DM}) by the corresponding DM uncertainty, that is,
\begin{equation}
\text{DM}_{chiral}\lesssim\epsilon_{\text{DM}},
\label{constraint1}
\end{equation}
which allows us to restrain the magnitude of the bi-isotropic chiral parameter as
\begin{equation}
		\tilde{\xi}_{c}\lesssim\left(1.82\times 10^{-12}\right)\left(\frac{\epsilon_{\text{DM}}}{\text{pc cm}^{-3}}\right)\left(\frac{\text{k pc}}{d}\right), \label{constraint_DMchiral}
		\end{equation}
where we have adopted the angular frequency associated to the \textit{centre frequency}, namely, $\omega=2\pi\times 148.9$ MHz.

Considering now the Faraday rotation, the same procedure can be performed using the LOFAR data for RMs \cite{LOFAR_RM}, from which we take the measurement uncertainties, here denoted by  $\epsilon_{\text{RM}}$. Similarly to the latter case, we consider $\epsilon_{\text{RM}}$ as the upper magnitude for the chiral contribution in (\ref{Faraday_chiral}), that is, $\text{RM}_{chiral}\lesssim\epsilon_{\text{RM}}$. In doing so, the RM constraints on the chiral parameter are determined by
\begin{equation}
	\tilde{\xi}_{c}\lesssim\left(4.2\times 10^{-20}\right)\left(\frac{\epsilon_{\text{RM}}}{\text{rad m}^{-2}}\right)\left(\frac{\text{k pc}}{d}\right), \label{constraint_RMchiral}
\end{equation}
where we take $\lambda\approx 2.01338$ m (associated with the centre frequency, $\nu=148.9~\text{MHz}$). As for the pulsar distance from the Earth, $d$, appearing in (\ref{constraint_DMchiral}) and (\ref{constraint_RMchiral}), we consider the \textit{corrected distances} listed in Ref.~\cite{Verbiest}, given in (k pc).

Taking into account the relation (\ref{constraint_DMchiral}) and the LOFAR data for DMs \cite{LOFAR}, {the upper constraints obtained for the magnitude of $\tilde{\xi}_{c}$ (dimensionless in natural units) are of} the order of $10^{-16}$ and $10^{-14}$, as presented in the fourth column in Table \ref{tab:table_constraints}. On the other hand, for the RM data \cite{LOFAR_RM} the relation (\ref{constraint_RMchiral}) yields constraints as tight as $10^{-22}$ and $10^{-21}$, as shown in the last column in Table \ref{tab:table_constraints}.

\begin{widetext}

	
	\begin{table}[htbb]
		\centering
		\caption{Constraints on the chiral parameter using pulsar DM and RM data.}
		\label{tab:table_constraints}
		
		\renewcommand{\arraystretch}{1.2}
		\setlength{\tabcolsep}{12pt}
		
		\begin{tabular}{*{6}{c}}
			\toprule[0.8pt]\midrule
			\textbf{Pulsars} & $d$ (k pc)$^{*}$ &  $\mathrm{DM}_{\mathrm{obs}}$ ($\mathrm{pc \,cm}^{-3}$) $^{\dagger}$ &  $\tilde{\xi}_{c}$  & $\mathrm{RM}_{\mathrm{obs}}$ {($\mathrm{rad~ m}^{-2}$)}	$^{\ddagger}$ & $\tilde{\xi}_{c}$ \\
			\midrule
			B1919+21 & $0.3$ &  $12.44399(63)$ &  $3.8\times 10^{-15}$ & $-15.04 \pm 0.02$  & $2.8\times 10^{-21}$  \\
			B1944+17 & $0.3$  & $16.1356(73)$     &   $4.4\times 10^{-14}$ & $-43.64 \pm 0.02$   & $2.8\times 10^{-21}$   \\
			
			B1929+10 & $0.31$ &   $3.18321(16)$   &    $9.3\times 10^{-16}$ &  $-5.27 \pm 0.01$ & $1.3\times 10^{-21}$   \\
			
			B2016+28  & $0.98$ &  $14.1839(13)$  &    $2.4\times 10^{-15}$ &  $-33.14 \pm 0.01$ &  $4.3\times 10^{-22}$    \\
			
			B2020+28  & $2.1$ & $24.63109(18)$ &   $1.5\times 10^{-16}$  & $-72.56\pm 0.02$  & $4.0\times 10^{-22}$     \\
			\midrule\bottomrule[0.8pt]
		\end{tabular}
		
	\end{table}
\begin{minipage}{0.8\linewidth} 
	\raggedright 
	\small{
		$^{*}$ {We have used the pulsar distances provided in Ref.~\cite{Verbiest}.} \\
		$^{\dagger}$ {We adopted LOFAR data for the dispersion measure \cite{LOFAR}.} \\
		$^{\ddagger}$ {We adopted LOFAR data for the rotation measure \cite{LOFAR_RM}.}
	}
\end{minipage}

\end{widetext}

\section{Final remarks \label{conclusion}}

In this work, we have investigated the propagation of electromagnetic waves in a {cold and magnetized} bi-isotropic-like chiral plasma {by discussing} optical properties and establishing upper bounds on the chiral parameter magnitude. {The bi-isotropic constitutive relations allowed the derivation of four} modified refractive indices given in Eqs. (\ref{n-R-M-indices-1_bi_isotropic}) and (\ref{n-L-E-indices-1_bi_isotropic}), associated with RCP and LCP propagating modes, respectively. {The chiral term promotes significant modifications to the electromagnetic propagation inside this plasma medium.} {For instance, in Sec. \ref{Propag.TL}, negative refraction behavior appears in the range $\omega_{c}<\omega<\omega_{-}$ for the index $n_{R_{+}}$ (see Fig.~\ref{NR+_bi-isotropic}),} with absorption ($\mathrm{Im}[n_{R_{+}}]\ne0$) occurring for $\omega_{c}<\omega<\omega_{R+}$ and free {(metamaterial-like)} propagation in the range $\omega_{R+}<\omega<\omega_{-}$. {The existence of negative refraction in a} chiral plasma {has already been} investigated and represents {a typical feature} of bi-isotropic plasmas \cite{Gao, Guo}. {Moreover, in the low-frequency regime, modified propagating RCP helicons arise exhibiting a linear dependence in the chiral parameter $\tilde{\xi_{c}}$, as discussed in Sec.~\ref{section-helicons}.}

{{By supposing a finite chiral parameter, we study} optical effects {expected to occur in} the cold plasma system, such as optical rotation and dichroism. In this sense, we have discussed its consequences {through both} the RP and dichroism coefficients, {which were determined starting from} the refractive indices $n_{R_{+}}$, $n_{L_{+}}$ and $n_{L_{-}}$ {(see Sec. \ref{birefringence})}.} We highlight that the RP $\delta_{LR}^{++}$ {defined by} \eqref{RP_RL+_bi}, depicted in Fig. \ref{Plot_RP2_bi}, presents a double sign reversal at the frequencies $\omega=\omega^{\prime}$ and $\omega=\omega^{\prime\prime}$, located in the range $0<\omega<\omega_{c}$. For $\omega> \omega_{c}$, the RP is negative and decreases with the frequency, as shown in Fig. \ref{Plot_RP1_bi}. {Likewise, the double reversal also appears in interval $0<\omega<\omega_{+}$ for the RP $\delta_{LR}^{-+}$ given in (\ref{RP_RL-+}), as displayed} in Fig. \ref{Plot_RP3_bi}. Although cold plasmas do not exhibit {such a double RP reversal, a} single reversion {already was reported in both} rotating plasmas \cite{Gueroult} and cold chiral plasma ruled by the axion electrodynamics \cite{Filipe1, Filipe2}. {Notwithstanding being a noncommon optical signature, the double RP reversal has also been encountered} in bi-isotropic media with anomalous Hall current \cite{Alex}. {This peculiar birefringent property} may represent a route for an optical characterization of the bi-isotropic chiral plasmas {in realistic experimental setups.}

{Similarly, Sec.~\ref{secDC} examines the effects of the chiral term on the dichroism phenomenon through coefficients $\delta_{dRL}^{++}$ and $\delta_{dLR}{-+}$, which} {are non-null only in the intervals $0<\omega<\omega_{R+}$ and $\omega_{c}<\omega<\omega_{R-}$,} {such as illustrated in} Fig.~\ref{Plot_DC1_bi}. {Compared to the standard case, the chiral parameter of bi-isotropic plasma narrows the zones where dichroism occurs, meaning that attenuation-free propagation {of larger frequency bands is} now allowed.}

In order to conciliate the hypothesis of a bi-isotropic-like plasma {permeating the interstellar} space with available observational data, {we have addressed} an ISM plasma model ruled by the Maxwell electrodynamics endowed with the constitutive relations (\ref{constitutiveRELATIONS2}), yielding the modified refractive {indices discussed in Sec. \ref{Propag.TL}. Thus, in Sec. \ref{Sec_Astrophysical_Aplication}, we have {utilized} astrophysical data} of dispersion and rotation measures {obtained from} radio signals traveling in a low-density ionized medium {to analyze the implications of the chiral parameter in the electromagnetic propagation. Such effects have allowed us to find upper bounds constraining the chiral variable.} From the time delay (\ref{timedelay_Chiral}), a chiral dispersion measure contribution, $\text{DM}_{chiral}$, was obtained and limited to uncertainty measures ($\epsilon_{\text{DM}}$) according to (\ref{constraint_DMchiral}). {Analogously,} a chiral rotation measure, $\text{RM}_{chiral}$, was obtained from the Faraday rotation (\ref{Faraday_chiral}), and limited to uncertainty $\epsilon_{\text{RM}}$ in (\ref{constraint_RMchiral}). {Similarly, we impose} astrophysical limits on the chiral parameter {by adopting} data of dispersion and rotation measures {attained for} five pulsars: B1919+21, B1944+17, B1929+10, B2016+28, and B2020+28. {Then, by using} dispersion measure (DM) data, the $\tilde{\xi}$ magnitude was restricted to the orders between $10^{-16}$ -- $10^{-18}$, as presented {in the fourth column of} Table \ref{tab:table_constraints}. {Furthermore,} {utilizing} rotation measure (RM) {information,} the chiral factor was restrained as $\tilde{\xi} \lesssim 10^{-22}$ -- $10^{-21}$ (see the last column in Table \ref{tab:table_constraints}).

	\begin{acknowledgments}
	
The authors express their gratitude to FAPEMA, CNPq, and CAPES (Brazilian research agencies) for their invaluable financial support. M.M.F. is supported by Grants No. FAPEMA APP-12151/22, No. CNPq/Produtividade 317048/2023-6, and No. CNPq/Universal/422527/2021-1. P.D.S.S. is grateful to FAPEMA APP-12151/22. R. C. acknowledges the support from the Grants No.  CNPq/312155/2023-9, No. FAPEMA/UNIVERSAL 00812/19, and No. FAPEMA APP-12299/22. Furthermore, we are indebted to CAPES/Finance Code 001 and FAPEMA/POS-GRAD-04755/24.

	\end{acknowledgments}

\section*{DATA AVAILABILITY}
The data that support the findings of this article are openly available in Refs. \cite{LOFAR,LOFAR_RM,Verbiest}.

\end{document}